\documentclass[a4paper,11pt,final]{article}

\usepackage{authblk}

\usepackage{graphics}
\usepackage{graphicx}
\usepackage{epsfig}

\usepackage{amsthm}

\usepackage{amsmath}
\usepackage{amssymb}

\date{}

\begin{document}

\title{Homogenization Approaches to Multiphase Lattice Random Walks}

\author[1]{Massimiliano Giona$^*$}
\author[1]{Davide Cocco}
\affil[1]{Dipartimento di Ingegneria Chimica DICMA
Facolt\`{a} di Ingegneria, La Sapienza Universit\`{a} di Roma
via Eudossiana 18, 00184, Roma, Italy  \authorcr
$^*$  Email: massimiliano.giona@uniroma1.it}

\maketitle

\begin{abstract}
This article analyzes several different homogenization
approaches to the long-term properties of multiphase
lattice random walks, recently introduced by Giona and Cocco
\cite{prl}, and characterized by  different  
values of  the hopping times and of 
 the distance between neighboring sites in each lattice phase.
Both parabolic and hyperbolic models are considered.
While all the parabolic models deriving from microscopic
 Langevin equations
driven by Wiener processes fail to predict the
long-term hydrodynamic behavior observed in lattice models, 
the discontinuous parabolic model,
in which the phase partition coefficient is {\em a-priori}
imposed, provides the correct answer. The implications
of this result as regards the connection
between equilibrium constraints and non-equilibrium
transport properties is thoroughly addressed. 
\end{abstract}

\section{Introduction}
\label{sec1}
Lattice Random Walks (LRW, for short) represent an invaluable source
of simple models and theoretical inspiration for assessing
the physics of interacting particle systems and for deriving,
from elementary and controllable microscopic rules for
particle motion and particle-particle interactions, macroscopic
hydrodynamic models \cite{gen1,gen2,gen3}. In the last
decades the physics of complex systems has achieved significant
advances thanks to the development of elementary lattice
models out of which explaining and deriving macroscopic
emergent features: the Ising, Glauber-Ising,   Kawasaki, damage-spreading
 models \cite{ising, kawasaki},
zero-range processes \cite{zrp,zrp1}, just to quote some of them introduced
for addressing phase-transitions and  condensation.

A central issue in the analysis of lattice models is the
derivation, from  simple rules defining lattice dynamics,
the macroscopic continuous hydrodynamic limit expressed
in terms of concentrations and fields defined in a continuous
space-time \cite{hydrolimit, hydrolimit1}.

Recently, by considering the simplest lattice model,
namely the lattice random walk for an ensemble of
independent particles, it has been shown that
a continuous hydrodynamic description
 is possible without imposing  the limit
of vanishing
 space- and time-scales.
This approach
leads to a hyperbolic continuous  transport model \cite{giona_lrw},
analogous to those derived in the framework of Generalized
Poisson-Kac processes \cite{gpk0,gpk1,gpk2,gpk3}.
The hyperbolic hydrodynamic model for asymmetric LRW
not only provides the correct scaling of the lower-order
moments with time (mean and square variance), but
accurately describes the early stages of the process,
when, starting e.g.  from an impulsive initial distribution, 
the probability density function is still  far away from
a Gaussian behavior. A further extension of this
approach is provided by the definition of Multiphase
Lattice Random Walk \cite{prl}. The Multiphase LRW, henceforth 
MuPh-LRW,
is a random walk on a multiphase lattice, characterized at a given
lattice point by the variation of the lattice-spacing and hopping time.
This setting, as discussed in \cite{prl}, determines the
occurrence of two distinct phases and, depending on the
lattice parameters, of a discontinuity in the probability
density function at the interface between the two lattice
phases.
MuPh-LRW provides a well-defined  physical lattice example for which
the use of the hyperbolic continuous hydrodynamic models
developed in \cite{giona_lrw} proves its validity with respect
to the parabolic counterparts, as it naturally permits
to identify, for ideal interfaces (see Section \ref{sec2}), the proper boundary
conditions to be set at the point of discontinuity (interface)
between the two lattice phases.
The natural development of this analysis is the study
of long-term dispersion properties in  a periodic structure
composed by the repetition of a unit cell in which two
distinct lattice phases are present.
This problem has been numerically approached in
 \cite{prl}. The lattice simulation
results are in perfect agreement with
the hyperbolic theory and, in some cases, cannot find
a correspondence in the long-term behavior of the
associated parabolic models based on Langevin equations
driven by Wiener fluctuations. The latter claim
is essentially based on the detailed analysis
of the long-term/large-distance properties of the hyperbolic
transport model for MuPh-LRW and of its parabolic counterparts.
The scope of this article is essentially to provide
the analytical background to this claim, based on
the homogenization theory of MuPh-LRW continuous models
grounded on moment analysis.
This analysis does not only present some novelty (especially
as regards the hyperbolic model), but also reveals some tricky
issues associated with regularity of transport parameters,
that are interesting {\em per  se}, and justifies the
content of the present article.
Moreover, the detailed analysis of  the long-term dispersion
properties deriving from hyperbolic and parabolic
transport models permits to clearly  appreciate  their
limitations and the relations between thermodynamic
equilibrium properties and non-equilibrium transport
parameters. 

The article is organized as follows. Section \ref{sec2}
provides the setting of the problem. Starting from the
classical LRW and its hyperbolic continuous description,
the concept of MuPh-LRW  introduced in \cite{prl} is
briefly reviewed, and the homogenization approach based on
moment analysis formalized.
Section \ref{sec3} addresses the homogenization of parabolic
models that can be defined  
in  an infinite structure represented by 
the periodic repetition of a multiphase unit cell. 
Essentially, two classes of parabolic models are considered.
To begin with, the classical model deriving from a Langevin-Wiener
 description of particle motion is considered, using a continuous
family of stochastic calculi ($\lambda$-integrals)
for describing the effect of the stochastic perturbation \cite{kloeden}.
In this case, $\lambda=0$ corresponds to the Ito formulation,
$\lambda=1/2$ returns the Stratonovich recipe, while $\lambda=1$
refers to the H\"anggi-Klimontovich interpretation.
The second class of parabolic models is a discontinuous model,
in which the equilibrium conditions at the interface, expressed
via a phase partition coefficient amongst the two phases,
are {\em a priori} given. Section \ref{sec4} addresses
in detail the homogenization calculations for the
hyperbolic model associated with MuPh-LRW, providing the
derivation of the expression for the effective diffusion
coefficient (dispersion coefficient) used in \cite{prl}.
Section \ref{sec5}  discusses the  results obtained using the various
parabolic approaches presented and their comparison
 with the long-term properties
of the hyperbolic model. Moreover,  
some implications of the theory, presented in the broader perspective
of the mutual relatioships between
 equilibrium properties and non-equilibrium  dynamics,
 are discussed.

\section{Setting of the problem}
\label{sec2}

A symmetric LRW on ${\mathbb Z}$ is specified, in the physical
space, once two parameters are given: a characteristic lengthscale
$\delta$, corresponding to the physical distance between nearest
neighboring sites, and a characteristic timescale $\tau$
representing the hopping time for performing a jump from a
site to one of its nearest neighbors. In the symmetric
case, no further parameters are needed, since the probabilities
of jumping to the two nearest neighboring sites from any initial
state are equal. Consequently, in  the  physical space-time,
the  particle dynamics is expressed by the evolution
equation
$x_{n+1}= x_n \pm \delta$, with probability $1/2$, and
$t_{n+1}=t_n+\tau$.

Next, suppose that a discontinuity is added into the model,
namely that a site, say $x=0$, is the boundary site separating
the left part of the lattice, in which the characteristic space-time
parameters are $\delta=\delta_1$, $\tau=\tau_1$, from the
right part where $\delta=\delta_2$ and $\tau=\tau_2$, supposing
that $|\delta_2-\delta_1|+|\tau_2-\tau_1| >0$.
The occurrence of different values of the lattice
 parameters $(\delta_h,\tau_h)$
in the two sublattices, $h=1,2$,
 determines statically the occurrence of
two lattice phases, separated by the interfacial point at $x=0$,
which, by definition, is the only site interacting directly with
sites of the two phases.
For this reason, this model has been  referred to 
as a Multiphase LRW (MuPh-LRW, for short).

If equal  probabilities characterize
the jump of a particle from the interfacial
site to the nearest neighbouring sites
of the two phases, the interface is referred to as {\em ideal}.
Deviations from this symmetric behavior determine a preferential
selection of one of the two phases induced by  the local interfacial dynamics.
This case is referred to as {\em non-ideal interfacial conditions}.

\subsection{Hyperbolic model for MuPh-LRW}
\label{sec2_1}

The case of ideal interfaces has been analyzed in \cite{prl},
and the main results can be summarized as follows:
\begin{itemize}
\item the hyperbolic transport model derived in \cite{giona_lrw}
for classical LRW describes accurately the qualitative and
quantitative
properties of MuPh-LRW. More precisely, if the
interface
is located at $x=0$, and indicating with $p_{\pm,h}(x,t)$
the partial probability waves in each phase $h=1,2$, 
the statistical properties of MuPh-LRW are described by the
hyperbolic system
\begin{equation}
\frac{\partial p_{\pm,h}(x,t)}{\partial t}= \mp b_h \, \frac{\partial
p_{\pm,h}(x,t)}{\partial x} \mp \lambda_h \, \left [p_{+,h}(x,t)-p_{-,h}(x,t)
\right ]
\label{eq2_1}
\end{equation}
where
\begin{equation}
b_h= \frac{ \delta_h}{\tau_h} \, , \qquad
\lambda_h= \frac{1}{\tau_h} \, , \qquad h=1,2
\label{eq2_2}
\end{equation}
and the subscript  $h$ labels the parameters associated with the $h$-lattice
phase;
\item $p_{\pm,h}(x,t)$ are defined in two disjoint subsets
of the lattice, say $p_{\pm,1}(x,t)$ for $x<0$ and $p_{\pm,2}(x,t)$
for $x>0$. The boundary conditions at the interface between
the two phases located at $x=0$, assuming ideal interfacial
conditions, are simply expressed, within the hyperbolic
model,
by enforcing the continuity of the partial fluxes
$b_h \, p_{\pm,h}(x,t)$ across the interface, i.e.,
\begin{equation}
 \left . b_2 \, p_{\pm,2}(x,t)  \right |_{x=0} =
\left . b_1 \, p_{\pm,1}(x,t)  \right |_{x=0} 
\label{eq2_3}
\end{equation}
Since the overall concentration $p_h(x,t)$, and the associated
flux $J_h(x,t)$ are expressed by
\begin{equation}
p_h(x,t)=p_{+,h}(x,t)+p_{-,h}(x,t) \, , \qquad
J_h(x,t)=b_h \, \left [ p_{+,h}(x,t)-p_{-,h}(x,t) \right ]
\label{eq2_4}
\end{equation}
$h=1,2$, eq. (\ref{eq2_3}) implies automatically
the continuity of the fluxes (or better to say of the normal component
of the flux) at the interface,
\begin{equation}
\left . J_2(x,t) \right |_{x=0} = \left . J_1(x,t) \right |_{x=0}
\label{eq2_5}
\end{equation}
which is a unavoidable consistency condition to ensure probability (mass)
conservation, and the boundary condition for the overall concentrations
\begin{equation}
\left . p_2(x,t) \right |_{x=0} = \left . \frac{b_1}{b_2} \,  
p_1(x,t) \right |_{x=0}
\label{eq2_6}
\end{equation}
\end{itemize}
Eq. (\ref{eq2_6}) implies the occurrence of a concentration
discontinuity at an ideal interface whenever $b_1 \neq b_2$.
Since the velocities $b_h$, $h=1,2$, entering eq.  (\ref{eq2_1}),
and expressed via eq. (\ref{eq2_2}) as a function of
the lattice parameters $\delta_h$ and $\tau_h$, are not ``native''
quantities
in a parabolic description of a LRW, in which the only dimensional
group of lattice parameters  controlling the diffusive dynamics
 is the ratio
$\delta_h^2/\tau_h$, eq. (\ref{eq2_6}) provides a radical
shift of paradigm as regards the continuous hydrodynamic
characterization of LRW. This has been analyzed in \cite{prl},
and we return to this issue in Section \ref{sec5}.

\begin{figure}[h]
\begin{center}
\includegraphics[width=12cm]{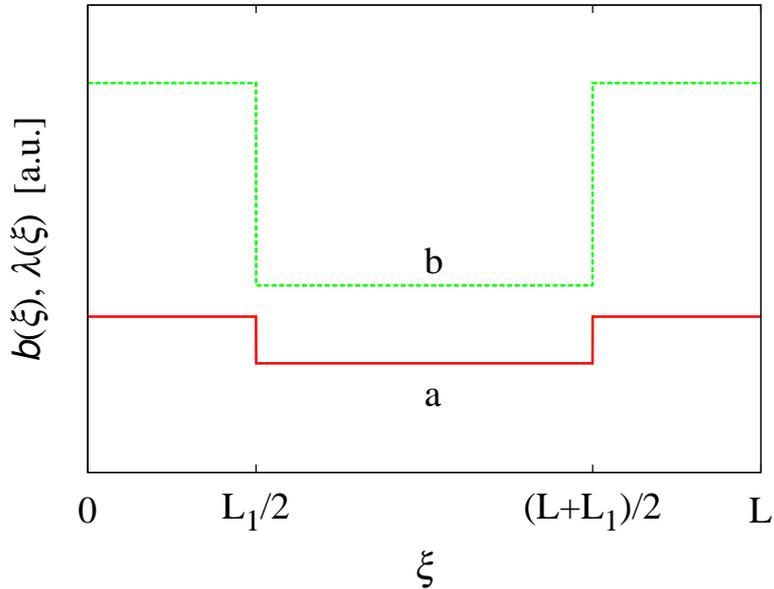}
\end{center}
\caption{Qualitative behavior of the fields $b(\xi)$ and $\lambda(\xi)$
vs $\xi$ within the periodicity cell. Line (a) refers to $b(\xi)$, while
line (b) to $\lambda(\xi)$.}
\label{Fig1}
\end{figure}

With reference to \cite{prl}, MuPh-LRW  has been studied
by considering its long-term/large-distance properties, in the case
particle motion occurs on a one-dimensional lattice constituted by
the periodic repetition of a unit lattice cell of length $L$, in which
a fraction $\phi_1=L_1/L$ of the cell is made of the lattice phase ``1'',
and the complementary part of phase ``2''.
The two phases are ordered, in the meaning that within
the periodicity cell only two interfacial points occurs.
Rephrasing this concept,
if we define with $b(x)$ and $\lambda(x)$ the velocities and transition
rates defining the hyperbolic model (\ref{eq2_1}), the spatial
behavior of these two quantities within the unit periodicity cell
of the lattice is qualitatively depicted in figure \ref{Fig1}.

It has been shown in \cite{prl} that a hyperbolic continuous
model  provides the accurate prediction
of the long-term dispersion properties observed in lattice
simulations of MuPh-LRW, and that the long-term behavior cannot be
explained by means of parabolic models associated with a Langevin
description of particle motion in the presence of Wiener fluctuations,
especially  whenever the lattice phase-heterogeneity involves
a discontinuity in the hopping times, i.e., $\tau_2 \neq \tau_1$.

In the remainder of this article, we develop the mathematical
details associated with the dispersion results presented in \cite{prl}.
Specifically, the closed-form calculations of the effective diffusion coefficient deriving from hyperbolic and parabolic models of MuPh-LRW 
discussed in \cite{prl} are presented in full length in Sections
\ref{sec3} and \ref{sec4}, respectively.
Moreover. an alternative parabolic model, referred to as the
``discontinuous parabolic model'', is analyzed, as it offers
 the opportunity  of 
imbedding the analysis of MuPh-LRW within the broader perspective of
the interplay between equilibrium properties and non-equilibrium
dynamics in relation to
the mathematical setting of the continuous hydrodynamic
model (see Section \ref{sec5}).

\subsection{Setting of homogenization analysis}
\label{sec2_2}

Let $p(x,t)$ be the probability density, solution of
a transport equation $\partial_ p(x,t)={\mathcal L}[p(x,t);x]$
in a periodic unbounded one-dimensional structure possesing period $L$,
i.e., ${\mathcal L}[f(x+L);x+L]={\mathcal L}[f(x);x]$,
for any periodic function $f(x)=f(x+L)$.
The space coordinate $x$ can be represented as a function
of a ``global'' integer coordinate $n \in {\mathbb Z}$,
indicating the unit cell  which $x$ refers  to,
and of a ``local'' coordinate $\xi \in (0,L)$ defining the
position within the unit cell, i.e.,
\begin{equation}
x= n  \, L + \xi
\label{eq2_2_0}
\end{equation}
so that $p(x,t)=p(n\, L + \xi,t)$.
Define the local moments of order $q=0,1,2,\dots$, as
\begin{equation}
p^{(q)}(\xi,t) =\sum_{n \in {\mathbb Z}} (n \, L + \xi)^q \, p(n \, L+\xi,t)
\label{eq2_2_1}
\end{equation}
The global $q$-order moment $M^{(q)}(t)$ 
of $p(x,t)$ can be expressed
as
\begin{equation}
M^{(q)}(t)= \int_{-\infty}^\infty x^n \, p(x,t) \, d x = \int_0^L
p^{(q)}(\xi,t) \, d \xi 
\label{eq2_2_2}
\end{equation}
i.e., as the integral with respect to the
 local coordinate $\xi$ inside the periodicity cell
of the local $q$-order moment $p^{(q)}(\xi,t)$.

The effective transport properties controlling the long-term/large-distance
evolution of $p(x,t)$, namely
 the effective velocity $V_{\rm eff}$ and the
effective diffusivity $D_{\rm eff}$, also referred to as the
dispersion coefficient, can be estimated from the long-term linear
scalings
\begin{eqnarray}
M^{(1)}(t) & = & V_{\rm eff} \, t + O(1) \nonumber \\
\sigma_x^2(t) & = & M^{(2)}(t)-\left [M^{(1)}(t) \right ]^2  = 
 2 \, D_{\rm eff} \, t + O(1)
\label{eq2_2_3}
\end{eqnarray}
where $O(1)$ indicates at most constant quantities.
The evaluation of $V_{\rm eff}$ and $D_{\rm eff}$ stems from the long-term
estimate of the dynamics of the lower-order local moments $p^{(q)}(\xi,t)$
that derives from the evolution equation for $p(x,t)$.

This is the classical approach to the homogenization theory in
periodic structures developed by Brenner and coworkers \cite{brenner_book}
and referred to as the ``macrotrasport paradigm'', originally deriving
from Aris analysis of solute dispersion in channel flows via moment analysis
\cite{aris}. In point of fact, there is a slight difference with
respect to the original Brenner approach, that uses
for the local moment the approximate expression
$p^{(q)}_{\rm Brenner}(\xi,t)=\sum_{n \in {\mathbb Z}} (n \, L)^q \, p(n \, L+ \xi,t)$,  valid solely in the long-time limit.
Conversely, eqs. (\ref{eq2_2_2}) is exact, as well as the
evolution equation for the local moments $p^{(q)}(\xi,t)$ 
that can be  derived from this position (see Sections \ref{sec3}
and \ref{sec4}). It can be also observed, that the
local $q$-order moments are periodic functions of $\xi$ of 
period $L$, while Brenner's moments do not fulfil this
property, and satisfy  a jump-boundary
conditions at the edges of the periodicity cell.
The periodicity of the local moments  simplifies
the homogenization analysis.

Henceforth, for all the models considered, be them parabolic or
hyperbolic,
we assume that the local transport parameters are
smooth and periodic functions of the position, and moreover 
that they
 are parametrized with respect to a small parameter $\varepsilon>0$,
such that, in the limit for  $\varepsilon \rightarrow 0$, the discontinuous
profile associated with the existence of the two lattice phases
within the unit cell is recovered.

To make an example, consider the parabolic models
deriving from a $\lambda$-integral interpretation of
the stochastic equation of motion of a particle in a periodic field
of diffusivity, representing a continuous stochastic approximation for
MuPh-LRW. In this case,
particle motion is described by a nonlinear Langevin-Wiener equation
\begin{equation}
d x(t) = \sqrt{2 \, D(x(t);\varepsilon)} *_\lambda d w(t)
\label{eq2_2_add1}
\end{equation}
where $D(x;\varepsilon)$  is a periodic function the position
with period $L$,
$D(x+L;\varepsilon)=D(x;\varepsilon)$,
that for any $\varepsilon>0$ is smooth, and for $\varepsilon$ tending to zero
\begin{equation}
\lim_{\varepsilon \rightarrow 0} D(x;\varepsilon)
= \left \{
\begin{array}{lll}
D_1 & & x \in (0,L_1) \\
D_2 & & x \in (L_1,L)
\end{array}
\right .
\label{eq2_2_5}
\end{equation}
where $L=L_1+L_2$, and $D_h$, $h=1,2$, are the diffusion
coefficients in the two lattice phases.
In eq. (\ref{eq2_2_add1}), $d w(t)$ are the increments in the
time interval $dt$ of a one-dimensional Wiener process and  the notation ``$*_\lambda$'' indicates 
the the stochastic Stieltjes integral over the increments of a Wiener
process is to be interpreted as a $\lambda$-integral \cite{kloeden}.
This means that given
$\lambda \in [0,1]$, and a function $f(w(t))$ of the realizations
of a Wiener process, the stochastic integral of $f(w)$
withe respect to the increments of the Wiener process over
the generic interval $[a,b]$ is given by
\begin{equation}
\int_{a}^b f(w(t)) *_\lambda d w(t) = \lim_{\delta_t \rightarrow 0}
\sum_{h=0}^{N-1} f \left ( (1-\lambda) w_h+\lambda w_{h+1}  \right )
\, (w_{h+1}-w_h)
\label{eq2_2_5add1}
\end{equation}
where $a=t_0<t_1 \cdots <t_N=b$, $w_h=w(t_h)$, and $\delta_t = 
\max_h (t_{h+1}-t_{h})$. For $\lambda=0,\,1/2,\,1$,  the
Ito,   Stratonovich  and  H\"anggi-Klimontovich formulation
of the stochastic integrals are respectively recovered.

The statistical characterization of the process involves
the probability density function $p(x,t)$, that is a solution
of the Fokker-Planck equation
\begin{equation}
\frac{\partial p(x,t)}{\partial t} = (1-\lambda) \, \frac{\partial }{\partial x}
\left [D^\prime(x;\varepsilon) \, p(x,t) \right ] +
\frac{\partial }{\partial x} \left [ D(x;\varepsilon) \, \frac{\partial
p(x,t)}{\partial x} \right ]
\label{eq2_2_4}
\end{equation}
where $D^\prime(x;\varepsilon)=d D(x;\varepsilon)/dx$ is also smooth for
$\varepsilon>0$.
An analogous approach applies to the transport parameters
entering the hyperbolic model.

Henceforth, for notational simplicity, the explicit dependence on 
$\varepsilon$ is eliminated, thus meaning
that $D(x)=D(x;\varepsilon)$, unless otherwise stated.

\section{Homogenization of parabolic models}
\label{sec3}

In this Section we consider the homogenization of the
parabolic  equations
describing in a continuous setting  particle motion
in multiphase lattices.

\subsection{$\lambda$-integral Fokker Planck equation}
\label{sec3_1}
Consider the Fokker-Planck equation for $p(x,t)$ in
the $\lambda$-integral meaning (\ref{eq2_2_4}).
Multiplying eq. (\ref{eq2_2_4}) by $(n \, L + \xi)^q$,
and summing over the  global integer coordinate $n$,
the evolution equation for $p^{(q)}(\xi,t)$, $\xi \in (0,L)$
is obtained
\begin{eqnarray}
\frac{\partial p^{(q)}(\xi,t)}{\partial t} & = & {\mathcal L}_{\xi}[p^{(q)}(\xi,t);\lambda]- q \, (1-\lambda) \, p^{(q-1)}(\xi,t) 
- q \frac{\partial \left [ D(\xi) \, p^{(q-1)}(\xi,t) \right ]}{\partial \xi}
\nonumber \\
& - & q \, D(\xi) \, \frac{\partial p^{(q-1)}(\xi,t)}{\partial \xi} 
+ q \, (q-1) \, D(\xi) \, p^{(q-2)}(\xi,t)
\label{eq3_1}
\end{eqnarray}
where we have used the property $\partial/\partial x=\partial/\partial \xi$
within each periodicity interval, and $ {\mathcal L}_{\xi}[\cdot;\lambda]$
indicates the  Fokker-Planck operator in the $\lambda$-representation
defined in the periodicity cell $\xi \in (0,L)$ by
\begin{equation}
 {\mathcal L}_{\xi}[p^{(q)}(\xi,t);\lambda]
= (1-\lambda) \, \frac{\partial \left [D^\prime(\xi) \, p^{(q)}(\xi,t) \right ]}
{\partial \xi}
+ \frac{\partial  }{\partial \xi} \left [ D(\xi) \, \frac{\partial p^{(q)}(\xi,t)}{\partial \xi } \right ]
\label{eq3_2}
\end{equation}
equipped with periodic boundary conditions,
\begin{equation}
p^{(q)}(0,t)=p^{(q)}(L,t) \, , \qquad
\left . \frac{\partial p^{(q)}(\xi,t)}{\partial \xi} \right |_{\xi=0}
= \left . \frac{\partial p^{(q)}(\xi,t)}{\partial \xi} \right |_{\xi=L}
\label{eq3_3}
\end{equation}
To begin with, consider the 0-th order moment $p^{(0)}(\xi,t)$,
solution of the equation $\partial_t p^{(0)}(\xi,t)={\mathcal L}_\xi[p^{(0)}(\xi,t);\lambda]$. In the long-term limit, $p^{(0)}(\xi,t)$ approaches
the stationary distribution $w_0(\xi)$ inside the periodicity
interval, solution of the equation ${\mathcal L}_\xi[w_{0}(\xi);\lambda]=0$
and given by
\begin{equation}
w_0(\xi)= \frac{A}{ D^{1-\lambda}(\xi) } \, , \qquad
A= \left [ \int_0^L \frac{ d \xi}{D^{1-\lambda}(\xi)}  \right ]^{-1}
\label{eq3_4}
\end{equation}
It follows that
\begin{equation}
w_0(\xi) \sim
\left \{
\begin{array}{lllll}
\frac{1}{D(\xi)} & & \lambda=0  & &(\mbox{Ito}) \\
\frac{1}{\sqrt{D(\xi)}} & & \lambda=1/2  & & (\mbox{Stratonovich}) \\
\mbox{const.} & & \lambda=1 & & (\mbox{H\"anggi-Klimontovich})
\end{array}
\right .
\label{eq3_5}
\end{equation}
For the first-order local moment $p^{(1)}(\xi,t)$,
eq. (\ref{eq3_1}) reduces to
\begin{equation}
\frac{\partial p^{(1)}(\xi,t)}{\partial t}= {\mathcal L}_\xi[p^{(1)}(\xi,t);\lambda] -(1-\lambda) \, D^\prime(\xi) \, p^{(0)}(\xi,t)
- \frac{\partial \left [ D(\xi) \, p^{(0)}(\xi,t) \right ] }{\partial \xi}
- D(\xi) \, \frac{\partial p^{(0)}(\xi,t)}{\partial \xi}
\label{eq3_6}
\end{equation}
In the long-term limit, $p^{(0)}(\xi,t) \rightarrow w_0(\xi)$.
From eq. (\ref{eq3_4}),  $w_0(\xi)$ is
 a function of $D(\xi)$, and  the periodicity of  both $D(\xi)$ 
and $p^{(q)}(\xi,t)$ implies  that
the integral of the r.h.s. of eq. (\ref{eq3_6}) over
the periodicity cell is vanishing. Thus,
\begin{equation}
\frac{d M^{(1)}(t)}{d t}= V_{\rm eff}= \int_0^L \frac{\partial p^{(1)}(\xi,t)}{\partial t}
\, d \xi =0
\label{eq3_7}
\end{equation}
meaning that the effective velocity is zero. 
Therefore,  in the long-term regime, $p^{(1)}(\xi,t)$ 
attains a stationary
profile $m_*^{(1)}(\xi)$, solution of the equation
\begin{equation}
{\mathcal L}_\xi[m_*^{(1)}(\xi);\lambda]= (1-\lambda) D^\prime(\xi)
\, w_0(\xi) + \frac{d \left [D(\xi) \, w_0(\xi) \right ]}{d \xi}+ D(\xi)
\, w_0^\prime(\xi) = \lambda \, A \, D^{\lambda-1}(\xi)  \, D^\prime(\xi)
\label{eq3_8}
\end{equation}
Integrating eq. (\ref{eq3_8}) with respect to $\xi$,
one obtains
\begin{equation}
D(\xi) \, \frac{d m_*^{(1)}(\xi)}{d \xi} + (1-\lambda) \, D^\prime(\xi)
\, m_*^{(1)}(\xi) = A \, D^\lambda(\xi) + C
\label{eq3_9}
\end{equation}
where $C$ is an integration constant, the value
of which follows by enforcing periodicity, i.e.,
$m_*^{(1)}(0)=m_*^{(1)}(L)$.
This leads to the expression for $m_*^{(1)}(\xi)$ in the
long-term regime
\begin{equation}
m_*^{(1)}(\xi)  = \frac{1}{D^{1-\lambda}(\xi)} \left [ A \, \xi + C \,
\int_0^\xi \frac{d \eta}{D^\lambda(\eta)}+ E  \right  ] 
\, , \qquad C = - A \, L \, \left [ \int_0^L \frac{d \xi}{D^\lambda(\xi)}
\right ]^{-1}
\label{eq3_10}
\end{equation}
where $E$ is an arbitrary integration constant, depending
on the initial conditions, the value of which, as shown below, does
not influence dispersion properties. 

Finally, the evolution of the second-order local moment
is defined by the equation
\begin{eqnarray}
\frac{\partial p^{(2)}(\xi,t)}{\partial t} &=& {\mathcal L}_\xi[p^{(2)}(\xi,t);\lambda] -2 \, (1-\lambda) \, D^\prime(\xi) \, p^{(1)}(\xi,t)
- 2 \, \frac{\partial \left [ D(\xi) \, p^{(1)}(\xi,t) \right ]}{\partial
\xi} \nonumber \\
&-& 2 \, D(\xi) \, \frac{\partial p^{(1)}(\xi,t)}{\partial \xi} 
+ 2 \, D(\xi) \, p^{(0)}(\xi,t) 
\label{eq3_11}
\end{eqnarray}
Since the effective velocity is vanishing, 
the integral of the second-order moment defines, modulo
an additive constant, the mean square displacement $\sigma_x^2(t)$,
and thus permits to estimate the effective diffusion coefficient $D_{\rm eff}$
\begin{equation}
\int_0^L p^{(2)}(\xi,t) \, d \xi = \frac{d \sigma_x^2(t)}{d t}= 2 \, D_{\rm eff}
\label{eq3_12}
\end{equation}
All the  factors expressed in divergence form, i.e., as 
spatial derivatives of a function, vanish because of periodicity,
so that the substitution of eq. (\ref{eq3_11})
into eq. (\ref{eq3_12}) in the long-term
regime, where $p^{(0)}(\xi,t) \rightarrow w_0(\xi)$,
$p^{(1)}(\xi,t) \rightarrow m_*^{(1)}(\xi)$, provides  the following expression
for $D_{\rm eff}$
\begin{equation}
D_{\rm eff}= \int_0^L D(\xi) \, w_0(\xi) \, d \xi- \lambda \,
\int_0^L D(\xi) \, \frac{d m_*^{(1)}(\xi)}{d \xi} \, d \xi
\label{eq3_13}
\end{equation}
that is the superposition of two contributions: (i) the average
of the position dependent diffusivity $D(\xi)$ with respect to
the stationary density $w_0(\xi)$, and a further contribution
depending on the derivative of $m_*^{(1)}(\xi)$. Due to
the functional structure of this second integral, the term
containing the arbitrary constant $E$ in eq. (\ref{eq3_10})
vanishes since $\int_0^L D(\xi) \left [d D^{-(2-\lambda)}(\xi)/d \xi
\right ] \, d \xi=0$ due to the periodicity of $D(\xi)$.

Upon an integration by parts, eq. (\ref{eq3_13}) can be expressed  also
as
\begin{equation}
D_{\rm eff}= \int_0^L D(\xi) \,  \, w_0(\xi) \, d \xi+ \lambda \,
\int_0^L m_*^{(1)}(\xi) \, D^\prime(\xi) \, d \xi
\label{eq3_13a1}
\end{equation}
The function $m_*^{(1)}(\xi)$ is smooth for any $\varepsilon>0$, and
in the limit for $\varepsilon \rightarrow 0$ it becomes
piecewise linear with discontinuities occurring at
the interfacial points separating the two lattice phases.
 Conversely, $D^\prime(\xi)$
approaches for $\varepsilon \rightarrow 0$ the superposition
of two Dirac's delta distributions of opposite amplitude
$\pm (D_2-D_1)$, centered at the interfacial points within
the periodicity cell. Apparently, the second integral
at the r.h.s of eq. (\ref{eq3_13a1}) is ill defined,
as the discontinuities of $m_*^{(1)}(\xi)$ occur exactly
at the interfacial points where the impulsive contributions
of $D^\prime(\xi)$ are centered. However, this is not
the case, for the reason that $m_*^{(1)}(\xi)$ is a functional of
$D(\xi)$ defined by eq. (\ref{eq3_10}),
 and eq. (\ref{eq3_13a1}) can be
further elaborated in order to obtain a more meaningful representation of 
$D_{\rm eff}$. Substituting into eq. (\ref{eq3_13a1}) the expressions
derived for $w_0(\xi)$, $A$, $C$ and $m_*^{(1)}(\xi)$, after some
quadraturae one arrives to the following compact expression for $D_{\rm eff}$
\begin{equation}
D_{\rm eff}= - L \, C = L^2 \, \left [\int_0^L \frac{d \xi}{D^{1-\lambda}(\xi)}
\right ]^{-1} \left [ \int_0^L \frac{d \xi}{D^\lambda(\xi)} \right ]^{-1}
\label{eq3_13a2}
\end{equation}
In the limit for $\varepsilon \rightarrow 0$, setting $\phi_h=L_h/L$,
eq. (\ref{eq3_13a2}) reduces to
\begin{equation}
D_{\rm eff}= \left ( \frac{\phi_1}{D_1^{1-\lambda}} +
\frac{\phi_2}{D_2^{1-\lambda}} \right )^{-1} \left ( \frac{\phi_1}{D_1^{\lambda}} +
\frac{\phi_2}{D_2^{\lambda}} \right )^{-1}
\label{eq3_13a3}
\end{equation}
that for $\phi_1=\phi_2=1/2$ simplifies as
\begin{equation}
\frac{4}{D_{\rm eff}} = \left  ( \frac{1}{D_1^{1-\lambda}} +
\frac{1}{D_2^{1-\lambda}} \right ) \, \left  ( \frac{1}{D_1^{\lambda}} +
\frac{1}{D_2^{\lambda}} \right ) 
\label{eq3_13a4}
\end{equation}

\subsection{Discontinuous parabolic model}
\label{sec3_2}

For further use, it is convenient to consider another
parabolic approximation not stemming from a stochastic
dynamics, but widely used in engineering applications,
namely a discontinuous parabolic model, where the
two lattice phases are kept distinct, possessing concentrations
$p_1(x,t)$ and $p_2(x,t)$, respectively, and
 satisfying the parabolic model
\begin{equation}
\frac{\partial p_h(x,t)}{\partial t}= D_h \, \frac{\partial^2 p_h(x,t)}{\partial
x^2} \; , \qquad x \in \Omega_h
\label{eq3_14}
\end{equation}
where $\Omega_h$ indicates the portion of the lattice composed by $h$-lattice
phase, $\Omega_1 \cup \Omega_2 = {\mathbb R}$.
Consequently, the support of each phase is the union of intervals
pertaining to each phase, and boundary conditions at phase interfaces
regulate probability partition amongst the phases.
Apart  from probability flux conservation,
\begin{equation}
\left . D_1 \, \frac{\partial p_1(x,t)}{\partial x} \right |_{\rm interface}
= \left . D_2  \, \frac{\partial p_2(x,t)}{\partial x} \right |_{\rm interface}  
\label{eq3_15}
\end{equation}
 assume a discontinuous
partition amongst the phases,
\begin{equation}
\left . p_2(x,t) \right |_{\rm interface} =  K \,
\left . p_1(x,t) \right |_{\rm interface}
\label{eq3_16}
\end{equation}
where $K>0$ is the  phase-partition coefficient. 
The physical origin of this model is discussed in paragraph \ref{sec5_3}.

So far, the
phase-partition coefficient is arbitrary, e.g. supposedly
known from empirical observations.
Also in this case, the local phase moments $p^{(q)}_h(\xi,t)= \sum_{n \in {\mathbb Z}} (n \, L + \xi)^q p_h(n \, L +\xi,t)$
can be defined. In the present case, it is convenient to
define the unit cell so that $\xi \in (0,L_1)$ corresponds to phase $1$ and
$\xi \in (L_1,L)$ to phase $2$, where $L=L_1+L_2$.
It is rather obvious that the local moments inherit the
boundary conditions (\ref{eq3_15})-(\ref{eq3_16}), so that for
any $q=0,1,	\dots$,
\begin{equation}
\left . p_2^{(q)}(\xi,t) \right |_{\xi=0,L_1} = K \, \left . p_1^{(q)}(\xi,t) \right |_{\xi=L,L_1}
\label{eq3_17}
\end{equation}
As regards the flux continuity, enforcing eq. (\ref{eq3_15}),
one obtains 
\begin{equation}
\left . D_1 \frac{\partial p_1^{(q)}(\xi,t)}{\partial \xi} - q \, D_1 \,
p_1^{(q-1)}(\xi,t) \right |_{\xi=0,L_1} =
\left . D_2 \frac{\partial p_2^{(q)}(\xi,t)}{\partial \xi} - q \, D_2 \,
p_2^{(q-1)}(\xi,t) \right |_{\xi=L,L_1}
\label{eq3_18}
\end{equation}
To begin with, consider the 0th-order local moments $p_h^{(0)}(\xi,t)$,
which satistfy the pure diffusion equation 
\begin{equation}
\frac{\partial p_h^{(0)}(\xi,t)}{\partial t}= D_h \, \frac{\partial p_h^{(0)}(\xi,t)}{\partial \xi^2}
\label{eq3_19}
\end{equation}
in their respective intervals of definition, i.e., $(0,L_1)$, and $(L_1,L)$,
equipped with the boundary conditions (\ref{eq3_17})-(\ref{eq3_18}) for
$q=0$.

In the long-term limit, the local 0th-order moments become stationary
$p_h^{(0)}(\xi,t) \rightarrow w_{h}^*(\xi)$,
and  uniform
within each interval interval of definition 
\begin{equation}
w_{1}^*(\xi)=\pi_1= \frac{1}{L_1+K \, L_2} \,  \; \; \xi \in (0,L_1) \, ,
\qquad
w_{2}^*(\xi)=\pi_2= \frac{K}{L_1+K \, L_2} \,  \; \; \xi \in (L_1,L)
\label{eq3_20}
\end{equation}
Next, consider the first-order local moments $p_h^{(1)}(\xi,t)$.
In each domain of definition, they satisfy the
equations
\begin{equation}
\frac{\partial p_h^{(1)}(\xi,t)}{\partial t}= D_h \, \frac{\partial^2 p_h^{(1)}(\xi,t)}{\partial \xi^2} - 2 \, D_h \, \frac{\partial p_h^{(0)}(\xi,t)}{\partial
\xi }
\label{eq3_21}
\end{equation}
In the long-term limit, $p_h^{(0)}(\xi,t)$ attain a uniform
distribution, so that the last term in eq. (\ref{eq3_21})
vanishes. Consequently,  $p_h^{(1)}(\xi,t) \rightarrow m_{h,*}^{(1)}(\xi)$ 
and the stationary $m_{h,*}^{(1)}(\xi)$ are linear functions of their
argument,
\begin{eqnarray}
m_{1,*}^{(1)}(\xi) &= & a + b \, \xi \, ,\qquad \xi \in (0,L_1) \nonumber \\
m_{2,*}^{(1)}(\xi) &= & c + d \, (\xi-L_1) \, ,\qquad \xi \in (L_2,L) 
\label{eq3_22}
\end{eqnarray}
where the constants $a,\,b,\,c,\,d$ should be
determined from the boundary conditions (\ref{eq3_17})-(\ref{eq3_18})
at $q=1$. 
Therefore, in the long-time limit the effective velocity
is identically vanishing, i.e., $V_{\rm eff}=0$.

From the boundary conditions one obtains three independent relations for
these constants,  
and one of
these   can be set equal to zero, say $a=0$.
The solution of the linear system for the remaining ones
provides
\begin{equation}
b = \frac{ L_2 \, (D_1 \, \pi_1 - D_2 \, \pi_2)}{\Delta}
\; , \qquad
c = K \, L_1 \, b \, , \qquad
d = -\frac{ K \, L_1 \, (D_1 \, \pi_1 - D_2 \, \pi_2)}{\Delta}
\label{eq3_23}
\end{equation}
where $\Delta= D_1 \, L_2 + K \, D_2 \, L_1$.
Finally, consider the second-order local moments $p_h^{(2)}(\xi,t)$
that satisfy the equations
\begin{equation}
\frac{\partial p_h^{(2)}(\xi,t)}{\partial t}=
D_h \, \frac{\partial^2 p_h^{(2)}(\xi,t)}{\partial \xi^2} - 4 \, D_h \,
\frac{\partial p_h^{(1)}(\xi,t)}{\partial \xi} + 2 \, D_h \, p_h^{(0)}(\xi,t)
\label{eq3_24}
\end{equation}
Since the effective velocity is vanishing, the time derivative of
the mean square displacement is simply expressed by
\begin{equation}
\frac{d \sigma_x^2(t)}{d t }= \int_0^{L_1} \frac{\partial p_1^{(2)}(\xi,t)}{\partial t} \, d \xi + \int_{L_1}^{L} \frac{\partial p_2^{(2)}(\xi,t)}{\partial t} \, d \xi 
\label{eq3_25}
\end{equation}
Making use of the balance equations 
for the local moments (\ref{eq3_25}), and enforcing the long-term
expression for the 0th-order moments (\ref{eq3_20})  one obtains 
\begin{eqnarray}
\frac{d \sigma_x^2(t)}{d t} &=  & \left (  \left . D_1 \frac{\partial p_1^{(2)}}
{\partial \xi} \right |_{\xi=L_1} - \left . D_2 \frac{\partial p_2^{(2)}}
{\partial \xi} \right |_{\xi=L_1} \right ) -
\left (  \left . D_1 \frac{\partial p_1^{(2)}}
{\partial \xi} \right |_{\xi=0} - \left . D_2 \frac{\partial p_2^{(2)}}
{\partial \xi} \right |_{\xi=L} \right ) \nonumber  \\
&- & 4 \left ( \left. D_1 \, p_1^{(1)}  \right |_{\xi=L_1}- \left . 
D_2 \, p_2^{(1)} \right  |_{\xi=L_1} \right ) + 4 \left ( \left. D_1 \, p_1^{(1)}  \right |_{\xi=0}- \left . 
D_2 \, p_2^{(1)} \right  |_{\xi=L}
\right )
 \nonumber \\
& + & 2 \left ( D_1 \, \pi_1 \, L_1 + D_2 \, \pi_2 \, L_2 \right )
\label{eq3_26}
\end{eqnarray}
Enforcing the boundary conditions for the second-order moments (\ref{eq3_18})
for $q=2$,  rearranging the order of the
various terms and enforcing the stationary profile of the
first-order local moments eq. (\ref{eq3_22}), eq. (\ref{eq3_26}) becomes
\begin{eqnarray}
\frac{d \sigma_x^2(t)}{d t}   &= &  2 \, \left ( D_1 \,  \pi_1 \, L_1 +
D_2 \, \pi_2 \, L_2 \right ) - 2 \, \left ( \left . D_1 \, p_1^{(1)} 
\right |_{\xi=L_1} - \left . D_2 \, p_2^{(1)} 
\right |_{\xi=L_1} \right )  \nonumber \\
&+& 2 \, \left ( \left . D_1 \, p_1^{(1)} 
\right |_{\xi=0} - \left . D_2 \, p_2^{(1)} 
\right |_{\xi=L} \right ) \nonumber \\
& = & 2 \, \left ( D_1 \,  \pi_1 \, L_1 +
D_2 \, \pi_2 \, L_2 \right )- 2 \, D_1 \, \left ( \left .  p_1^{(1)} 
\right |_{\xi=L_1}-\left .  p_1^{(1)} 
\right |_{\xi=0} \right ) \nonumber \\
& - & 2 \, D_2 \, 
\left ( \left .  p_2^{(1)} 
\right |_{\xi=L}-\left .  p_2^{(1)} 
\right |_{\xi=L_1} \right ) \nonumber \\
& = & 2 \, \left ( D_1 \,  \pi_1 \, L_1 +
D_2 \, \pi_2 \, L_2 \right ) - 2 \, D_1 \, L_1  \, b - 2 \, D_2 \, L_2 \, d
\label{eq3_27}
\end{eqnarray}
where $b$ and $d$ are the slopes of the linear behavior of
$m_{h,*}^{(1)}(\xi)$ with $\xi$ in the respective intervals
of definition. Observe that eq. (\ref{eq3_27}) depends solely
on the slopes of the first-order local moments, and this
justifies why the value of coefficient $a$ in  (\ref{eq3_22})
is absolutely irrelevant as regards the dispersion properties.
From eq. (\ref{eq3_27}), substituting the values  for $b$ and $d$,
eq. (\ref{eq3_23}), the  expression for the long-term
dispersion coefficient follows
\begin{equation}
D_{\rm eff}= \left ( D_1 \, \pi_1 \, L_1 +D_2 \, \pi_2 \, L_2 \right )
- \frac{L_1 \, L_2}{D_1 \, L_2 + K \, D_2 \, L_1}
\, \left (D_1 - K \, D_2 \right ) \, \left (D_1 \, \pi_1 -D_2 \, \pi_2
\right )
\label{eq3_28}
\end{equation}
In the particular case $L_1=L_2$, eq. (\ref{eq3_28}) attains
the simple and compact expression
\begin{equation}
D_{\rm eff}= \frac{4 \, D_1 \, D_2}{(1+K) \, (D_1+K \, D_2)}
\label{eq3_29}
\end{equation}

\section{Homogenization of the hyperbolic model}
\label{sec4}
In this Section, we consider the  homogenization of the
hyperbolic model  for MuPh-LRW 
in the presence of an ideal interface between the
two lattice phases.
As in the previous Section, we consider a family
of transport parameters, that in the case of the hyperbolic model
are the velocity $b(x;\varepsilon)$ and the transition rate $\lambda(x;\varepsilon)$, that are smooth functions of the position for $\varepsilon>0$,
periodic with period $L$
and that, in the limit of $\varepsilon \rightarrow 0$, converge
to the corresponding properties of the two lattice
phases,
\begin{equation}
\lim_{\varepsilon \rightarrow 0}  b(x;\varepsilon) =
\left \{
\begin{array}{lll}
b_1 & & x \in \Omega_1 \\
b_2 & & x \in \Omega_2 
\end{array}
\right .
\; , \qquad
\lim_{\varepsilon \rightarrow 0}  \lambda(x;\varepsilon) =
\left \{
\begin{array}{lll}
\lambda_1 & & x \in \Omega_1 \\
\lambda_2 & & x \in \Omega_2 
\end{array}
\right .
\label{eq4_1}
\end{equation}
As in the previous Section, we omit the explicit dependence on the
parameter $\varepsilon$ for notational convenience.
Therefore, the evolution equation for the
partial waves $p_\pm(x,t)$ associated with this model 
reads
\begin{equation}
\frac{\partial p_\pm(x,t)}{\partial t} = \mp \frac{\partial
\left [ b(x) \, p_\pm(x,t) \right ]}{\partial x} \mp \lambda(x)
\, \left [p_+(x,t) -p_-(x,t) \right ]
\label{eq4_2}
\end{equation}
Introducing the partial local moments of order $q$
\begin{equation}
p_\pm^{(q)}(\xi,t) = \sum_{n \in {\mathbb Z}} ( n \, L + \xi)^q
\, p_\pm(n \, L + \xi,t) 
\label{eq4_3}
\end{equation}
the overall global moments $M^{(q)}(t)$ of order $q$ are
expressed by
\begin{equation}
M^{(q)}(t)= \sum_{\alpha=\pm} \int_0^L p_\alpha^{(q)}(\xi,t)
\label{eq4_4}
\end{equation}
By definition, the partial local moments are periodic functions
of the local coordinate $\xi$
\begin{equation}
p_\pm^{(q)}(0,t)=p_\pm^{(q)}(L,t)
\label{eq4_4bis}
\end{equation}
and satisfy the balance equations
\begin{eqnarray}
\frac{\partial p_+^{(q)}(\xi,t)}{\partial t} & = &
- \frac{\partial \left [ b(\xi) \, p_+^{(q)}(\xi,t) \right ]}{\partial \xi}
+ q \, b(\xi) \, p_+^{(q-1)}(\xi,t) -\lambda(\xi) \, \left [ p_+^{(q)}(\xi,t)-
p_-^{(q)}(\xi,t) \right ] \nonumber \\
\frac{\partial p_-^{(q)}(\xi,t)}{\partial t} & = &
 \frac{\partial \left [ b(\xi) \, p_-^{(q)}(\xi,t) \right ]}{\partial \xi}
- q \, b(\xi) \, p_-^{(q-1)}(\xi,t) +\lambda(\xi) \, \left [ p_+^{(q)}(\xi,t)-
p_-^{(q)}(\xi,t) \right ]
\label{eq4_5}
\end{eqnarray}
The 0th order partial moments converge, in the long-time limit, to
the stationary equilibrium distributions $w_{0,\pm}(\xi)$,
solutions of the equations
\begin{equation}
\frac{d \left [ b(\xi) \, w_{0,\pm}(\xi) \right ]}{d \xi}= - \lambda(\xi)
\, \left [ w_{0,+}(\xi) - w_{0,-}(\xi) \right ]
\label{eq4_6}
\end{equation}
from which  it follows that
\begin{equation}
b(\xi) \, \left [w_{0,+}(\xi) - w_{0,-}(\xi) \right ] = C_0
\label{eq4_7}
\end{equation}
where $C_0$ is an integration constant that should be identically
vanishing because of periodicity $w_{0,\pm}(0)=w_{0,\pm}(L)$.
Consequently,
\begin{equation}
w_{0,+}(\xi)=w_{0,-}(\xi)=\frac{w_0(\xi)}{2}
\label{eq4_8}
\end{equation}
where $w_0(\xi)$ is given by
\begin{equation}
w_0(\xi)= \frac{A}{b(\xi)} \; , \qquad
A= \left [ \int_0^L \frac{d \xi}{b(\xi)} \right ]^{-1}
\label{eq4_9}
\end{equation}
Next, consider the first-order partial local moments satisfying
eq. (\ref{eq4_5}) with $q=1$. Integrating their balance equations
over the periodicity cell and summing the $\pm$-contributions
one obtains,
\begin{equation}
\frac{d M^{(1)}(t)}{d t}= \int_0^L b(\xi) \, \left [ p_+^{(0)}(\xi,t)-
p_-^{(0)}(\xi,t) \right ] \, d \xi
\label{eq4_10}
\end{equation}
Since in the long-time limit the two 0th order local
partial moments are equal to each other,
$d M^{(1)}(t){d t}=0$, and $V_{\rm eff}=0$.
In the long-time limit, the first-order  local partial 
moments attain a stationary profile $p_\pm^{(1)}(\xi,t) \rightarrow
m_{*,\pm}^{(1)}(\xi)$, solution of the stationary equations
(\ref{eq4_5}) with $p^{(0)}_\pm(\xi,t)$ substituted by
$w_{0,\pm}(\xi)=w_0(\xi)/2$. Also for the first-order moments
a relation analogous to eq. (\ref{eq4_7}) holds
\begin{equation}
b(\xi) \, \left [ m_{*,+}^{(1)}(\xi) - m_{*,-}^{(1)}(\xi) \right ] = C_1
\label{eq4_11}
\end{equation}
but the integration constant $C_1$ is not vanishing.
 In point of fact, making use of eq. (\ref{eq4_11})
within the balance equation   (\ref{eq4_5}) for $q=1$ for $p_+^{(1)}(\xi,t)=m_{*,+}^{(1)}(\xi)$ at
steady state, it follows that
\begin{equation}
\frac{d \left [ b(\xi) \, m_{*,+}^{(1)}(\xi) \right ]}{d \xi}= \frac{b(\xi) \, w_0(\xi)}{2} - \frac{C_1 \, \lambda(\xi)}{b(\xi)}
\label{eq4_12}
\end{equation}
and enforcing periodicity, $b(0)  \,
 m_{*,+}^{(1)}(0)= b(L)  \,  m_{*,+}^{(1)}(L)$,
one finally gets
\begin{equation}
\frac{1}{2} \int_0^L b(\xi) \, w_0(\xi) \, d \xi- C_1 \, \int_0^L \frac{\lambda(\xi)}{b(\xi)} \, d \xi =0
\label{eq4_13}
\end{equation}
that yields for $C_1$
\begin{equation}
C_1 = \frac{\frac{1}{2} \int_0^L b(\xi) \, w_0(\xi) \, d \xi}{\int_0^L
\frac{\lambda(\xi)}{b(\xi)} \, d \xi }
\label{eq4_14}
\end{equation}
The expression for the dispersion coefficient $D_{\rm eff}$
is a direct consequence of eqs. (\ref{eq4_12}), (\ref{eq4_14}).
Integrating the balance equations for the second-order local
partial moments,  eq. (\ref{eq4_5}) with $q=2$,  over the
periodicity cell, summing with respect to $\pm$, and enforcing both
periodicity and the long-term behavior of $p_\pm^{(1)}(\xi,t)$,
it follows that
\begin{equation}
\frac{d \sigma_x^2(t)}{d t}=\frac{d M^{(2)}(t)}{d t}=
2 \,\int_0^L b(\xi) \, \left [m_{*,+}^{(1)}(\xi) - m_{*,-}^{(1)}(\xi) \right ]
\, d \xi = 2 \, C_1 \, L
\label{eq4_15}
\end{equation}
where the property of vanishing effective
velocity $V_{\rm eff}=0$ has been used.
It follows from eq. (\ref{eq4_15}) 
the expression for $D_{\rm eff}$
\begin{equation}
D_{\rm eff}= \frac{L}{2} \frac{\int_0^L b(\xi) \, w_0(\xi) \, d \xi}
{\int_0^L \frac{\lambda(\xi)}{b(\xi)} \, d \xi}
\label{eq4_16}
\end{equation}
In the limit for $\varepsilon \rightarrow 0$,
\begin{equation}
w_0(\xi) =
\left \{
\begin{array}{lll}
A/b_1 & & \xi \in (0,L_1) \\
A/b_2 & & \xi \in (L_2,L)
\end{array}
\right .
\label{eq4_17}
\end{equation}
and
\begin{equation}
\int_0^L b(\xi) \, w_0(\xi) \, d \xi = A \, L = L \, \left ( \frac{L_1}{b_1}
+ \frac{L_2}{b_2} \right )^{-1}
\label{eq4_18}
\end{equation}
and analogously
\begin{equation}
\int_0^L \frac{\lambda(\xi)}{b(\xi)} \, d \xi = \frac{L_1 \, \lambda_1}{b_1}
+ \frac{L_2 \, \lambda_2}{b_2}
\label{eq4_19}
\end{equation}
By considering that in the present formulation of the hyperbolic
model, the transition rates $\lambda_h$, $h=1,2$, are
related to the hopping times $\tau_h$ of the LRW by the
relation $\lambda_h=1/\tau_h$, the effective diffusion
coefficient, in the limit for $\varepsilon \rightarrow 0$,
corresponding to the occurrence of two distinct lattice phases,
can be expressed by
\begin{equation}
D_{\rm eff}= \frac{1}{2} \, \left ( \frac{\phi_1}{b_1}+ \frac{\phi_2}{b_2}
\right ) ^{-1} \left ( \frac{\phi_1}{b_1 \, \tau_1} + \frac{\phi_2}{b_2 \,
\tau_2} \right )^{-1}
\label{eq4_20}
\end{equation}
where $\phi_h=L_h/L$, $h=1,2$ are the fraction occupied
by the two lattice phases.
In the symmetric case $\phi_1=\phi_2=1/2$, eq. (\ref{eq4_20})
can be rewritten as
\begin{equation}
\frac{2}{D_{\rm eff}} = \left (\frac{1}{b_1} + \frac{1}{b_2} \right )
\, \left ( \frac{1}{b_1 \, \tau_1} + \frac{1}{b_2 \, \tau_2} \right )
\label{eq4_21}
\end{equation}

\section{Further observations}
\label{sec5}
In this Section we address some complementary/numerical issues
associated with the homogenization theory developed
in the previous two Sections. The analysis makes use
of the result shown in \cite{prl} that the hyperbolic
model provides the correct result for the effective
diffusion coefficient observed in MuPh-LWR model.

\subsection{Langevin-Ito dispersion}
\label{sec5_1}

\begin{figure}
\begin{center}
\includegraphics[width=12cm]{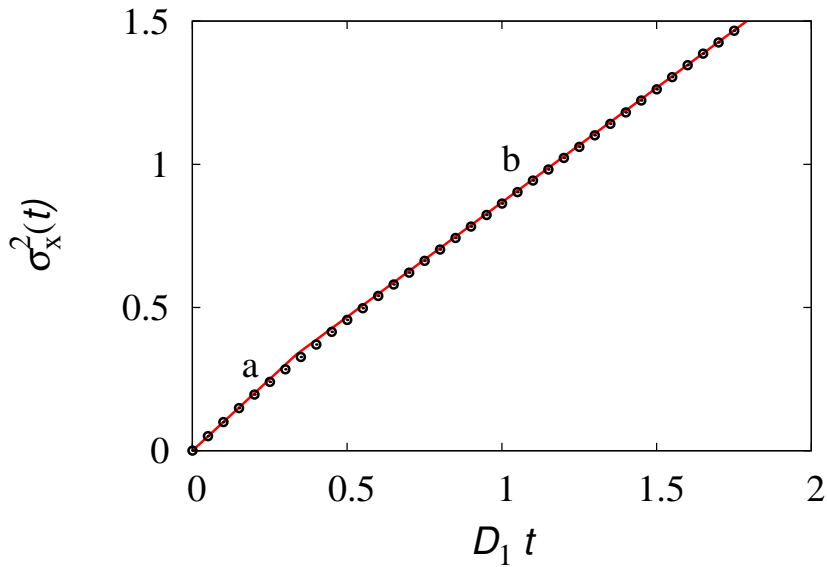}
\end{center}
\caption{Mean square displacement $\sigma_x^2(t)$ vs $D_1 \, t$ for the
Langevin-Ito approximation of MuPh-LRW at $D_2/D_1=1/4$, obtained
from stochastic simulations (symbols $\circ$)
 of eq. (\ref{eq2_2_add1}) with $\lambda=0$.
Line (a) refers to the initial linear scaling $\sigma_x^2(t) \sim
2 \, D_{\rm in} \, t$, where $D_{\rm min}$ is given by eq. (\ref{eq5_2}),
line (b) to the long-term scaling controlled by the
effective diffusivity $D_{\rm eff}$, eq. (\ref{eq3_13a4}) with
$\lambda=0$.}
\label{Fig2}
\end{figure}

To begin with, consider the long-term properties of the
Langevin-Ito equation (\ref{eq2_2_add1}) in  ${\mathbb R}$,
in the presence of a periodic diffusion coefficient, mimicking
the occurrence of two lattice phases $D(x+L)=D(x)$,
where
$D(\xi)=D_1$  for $\xi \in (0,L_1)$ and $D(\xi)=D_2$ for
$\xi \in (L_1,L)$. Set $L_1=L_2=L/2=1$. Figure
\ref{Fig2} depicts the behavior of the mean square
displacement $\sigma_x^2(t)$ as a function of time $t$
obtained from stochastic simulations of eq. (\ref{eq2_2_add1})
using an ensemble of $N_p=10^6$ particles initially located
at $x=0$, for $D_1=5 \times 10^{-5}$ and $D_2=D_1/4$.

\begin{figure}
\begin{center}
\includegraphics[width=12cm]{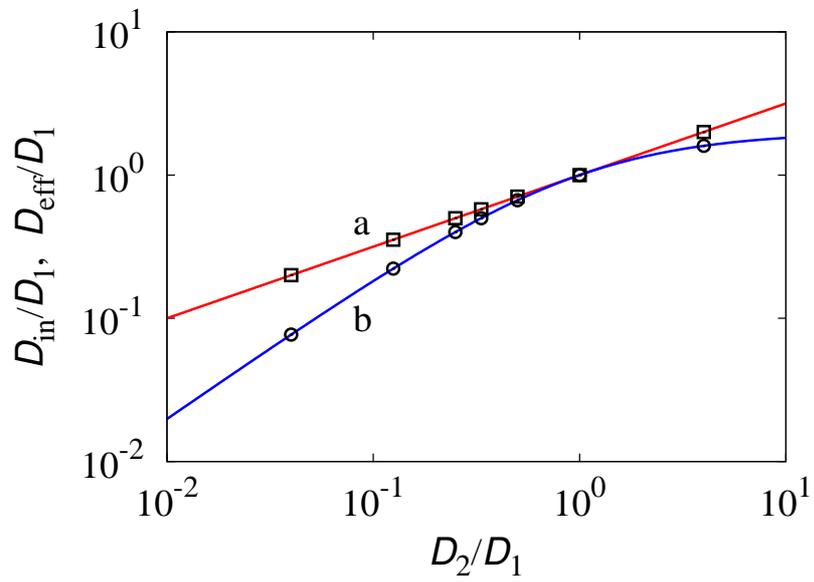}
\end{center}
\caption{$D_{\rm in}/D_1$ and $D_{\rm eff}/D_1$ vs $D_2/D_1$
for the Langevin-Ito approximation of MuPh-LRW.
Symbols $(\square, \, \circ) $ refer to stochastic simulation
results, lines (a) and (b) to the  empirical rule
(\ref{eq5_2}) and to the prediction of homogenization theory,
respectively.
Line (a) and $(\square)$ refer to $D_{\rm in}/D_1$,
line (b) and $(\circ)$ to $D_{\rm eff}/D_1$.
}
\label{Fig4}
\end{figure}

It can be observed that $\sigma_x^2(t)$ displays a crossover
from an initial linear scaling $\sigma_x^2(t)= 2 \, D_{\rm in} \, t$,
to the long-term behavior $\sigma_x^2(t) = 2 \, D_{\rm eff} \, t$.
The long-term effective diffusivity $D_{\rm eff}$ estimated from
stochastic simulations agrees with the homogenization prediction
(\ref{eq3_13a1}) or (\ref{eq3_13a4}), as shown in figure \ref{Fig4}.

The long-term diffusion coefficient $D_{\rm eff}$ in the
Langevin-Ito case corresponds to the average of the local
diffusivity $D(\xi)$ with respect to
the ergodic cell density $w_0(\xi)$, and eq. (\ref{eq3_13a1})
can be equivalent expressed as
\begin{equation}
D_{\rm eff}= \sum_{h=1}^2 D_h \, w_h
\label{eq5_1}
\end{equation}
where $w_h=A/D_h$, represent the fraction of time spent in
the $h$-th lattice phase. It should be observed that the
Langevin-Ito dynamics (i.e., $\lambda=0$) is the unique case in which this
representation of the effective diffusivity applies, as
for any $\lambda \neq 0$ the second term in eq. (\ref{eq3_13a1})
plays a crucial role in determining $D_{\rm eff}$, leading to
eq.  (\ref{eq3_13a4}).

The short-term scaling can be interpreted analogously,
as the average of the phase diffusivities with respect to
the short-time phase distribution $w_h^{(\rm in)}$, $h=1,2$,
that from numerical simulations  can be approximated
by the square-root expression $w_h^{({\rm in})}= B/\sqrt{D_h}$,
and $B$ is the normalization constant. It
follows from this observation that the short-term diffusivity
attains the approximate expression,
\begin{equation}
D_{\rm in}= \sqrt{D_1 \, D_2}
\label{eq5_2}
\end{equation}
which is just the geometric mean of the phase diffusivities
$D_h$. The quantitative agreement between
eq. (\ref{eq5_2}) and simulation results in depicted in 
figure \ref{Fig4} line (a).

\begin{figure}
\begin{center}
\includegraphics[width=12cm]{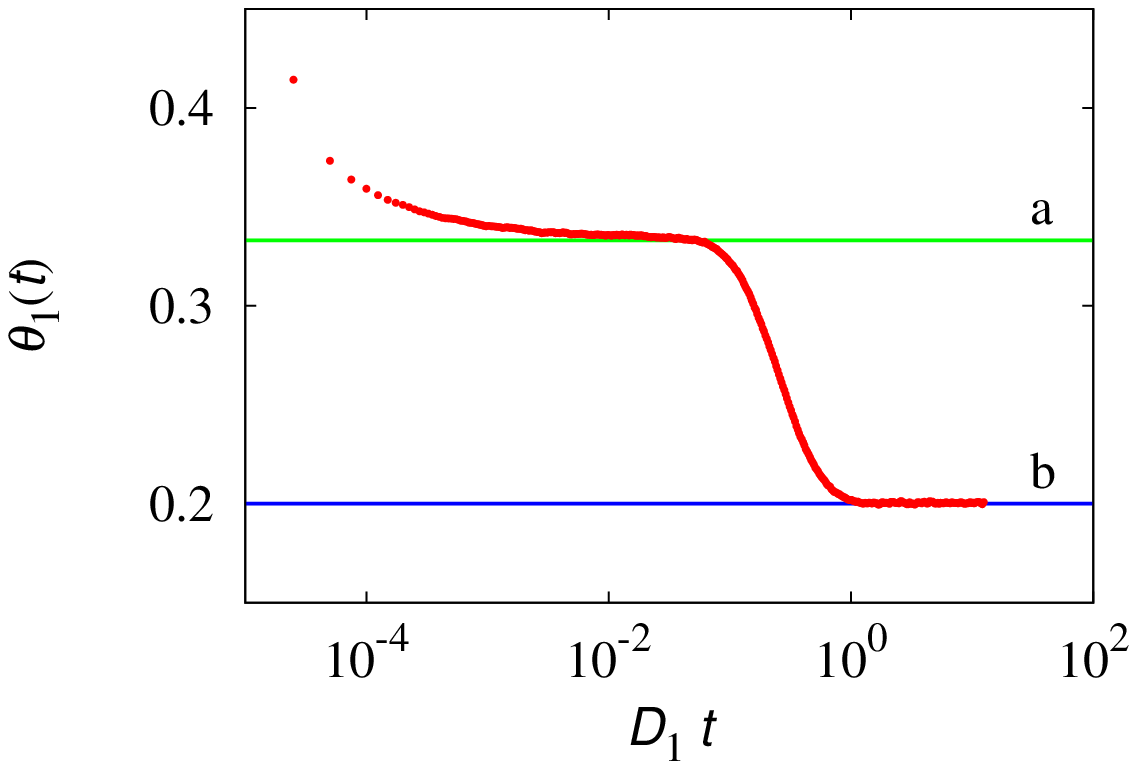}
\end{center}
\caption{Particle fraction in lattice phase ``1'' $\theta_1(t)$ vs $D_1 \, t$
for the Langevin-Ito approximation of MuPh-LRW, for $D_2/D_1=1/4$ obtained
from stochastic simulations. The horizontal lines (a) and (b)
represent $\theta_1(t)=1/3$, and $\theta_1(t)=\theta_{\rm equil}=1/5$.}
\label{Fig4bis}
\end{figure}

In point of fact, the interpretation of the
two short- and long-term diffusivities $D_{\rm in}$ and $D_{\rm eff}$
as the averages of the phase diffusivities with respect
to the the time-fractions spent by moving particles
in the two phases follows from the direct estimate of
these quantities. This phenomenon
is depicted in figure \ref{Fig4bis} that
shows the fraction $\theta_1(t)$ of particles located within
 phase ``1'' at time $t$, using a larger $N_p=10^8$ ensemble
of particles initially located at $x=0$, i.e, at an interface point.
The values of $D_h$ are the same as for figure \ref{Fig2}.

It can be observed, that at short timescales, $\theta_1(t)$
approaches an apparently constant value $\theta_1 \simeq 1/3$,
 at intermediate times $D_1 \, t \leq 0.2$,
that corresponds to $w_1^{({\rm in})}/w_2^{({\rm in})} = \sqrt{D_2/D_1} = 1/2$,
collapsing for $D_1 \, t \geq 1$ to the
equilibrium value $w_1/w_2=D_2/D_1=1/4$.

An interesting property of the Langevin-Ito model
stems from the comparison of eq. (\ref{eq3_13a4})
with eq. (\ref{eq4_21}) deriving from the hyperbolic transport model
that provides the correct expression found in lattice simulations
of MuPh-LRW \cite{prl}.
Assume that the characteristic length
of the two lattice phases are equal, i.e., $\delta_1=\delta_2=\delta$,
so that heterogeneity stems exclusively from the hopping
times $\tau_1 \neq \tau_2$, and set $\phi_1=\phi_2$.
The Langevin-Ito and the Langevin-H\"anggi-Klimontovich 
results for the effective diffusion coefficient are equal
(as eq. (\ref{eq3_13a4}) is invariant with respect to the
transformation $\lambda \mapsto 1-\lambda$), and 
simplifies as
\begin{equation}
\frac{1}{D_{\rm eff}} = \frac{1}{2} \left ( \frac{1}{D_1} + \frac{1}{D_2} 
\right )
\label{eqw_1}
\end{equation}
Next consider the expression deriving from the hyperbolic
hydrodynamic model, eq. (\ref{eq4_21}).
In this case $D_1=\delta^2/2 \, \tau_1$, $D_2=\delta^2/2 \, \tau_2$
and eq. ((\ref{eq4_21}) can be rewritten as
\begin{eqnarray}
\frac{1}{D_{\rm eff}}  & = & \frac{1}{2} \left (
\frac {\tau_1}{\delta} + \frac{\tau_2}{\delta} \right ) \, \frac{2}{\delta}
= \frac{\tau_1}{\delta^2} + \frac{\tau_2}{\delta^2}
\nonumber \\
& = & \frac{1}{2} \left ( \frac{1}{D_1} + \frac{1}{D_2} \right )
\label{eqw_2}
\end{eqnarray}
which coincides with the Langevin-Ito result (\ref{eqw_2}).

\subsection{Continuous hyperbolic models and  the Stratonovich limit}
\label{sec5_2}

The Langevin-Ito model discussed in the previous paragraph is
an interesting example of application of homogenization theory,
it describes correctly the long-term properties
if $\delta_1=\delta_2$, but fails in the case
the lattice spacing of the two phases are different.
A complementary situation is provides by the
Stratonovich approximation, that fails for $\delta_1 = \delta_2$,
but provides the correct answer for equal hopping times,
i.e., if $\tau_1=\tau_2$.

This is a consequence of  the theory of hyperbolic transport models
\cite{gpk1,gpk2,gpk3}. In the case the transition rates
are uniform,  i.e., $\lambda(x)=\lambda_0$ does not depend on $x$,
the hyperbolic model (\ref{eq4_2}), converges in the Kac limit
to the parabolic Fokker-Planck equation associated with
the Langevin dynamic (\ref{eq2_2_add1}), with $D(x)=b^2(x)/2 \lambda_0$
interpreted {\em a la } Stratonovich, and moreover their
long-term properties also coincide.

This result, 
 in the case $\tau_2=\tau_1=\tau=1/\lambda_0$,
follows straightforwardly from the comparison of eq. (\ref{eq4_21})
with (\ref{eq3_13a4}). Since $b_h= \delta_h/\tau$,
$D_h=\delta_h^2/2\, \tau$, expressing the velocities
$b_h$ entering  of eq. (\ref{eq4_21}) in terms of the
corresponding phase diffusivities $D_h$, eq.  (\ref{eq4_21})
provides 
\begin{equation}
\frac{2}{D_{\rm eff}} = \frac{1}{2} \left ( \frac{1}{\sqrt{D_1}} +
\frac{1}{\sqrt{D_2}} \right )^2
\label{eq5_3}
\end{equation}
that coincides with eq.  (\ref{eq3_13a4}) in the Stratonovich meaning.

\begin{figure}
\begin{center}
\includegraphics[width=16cm]{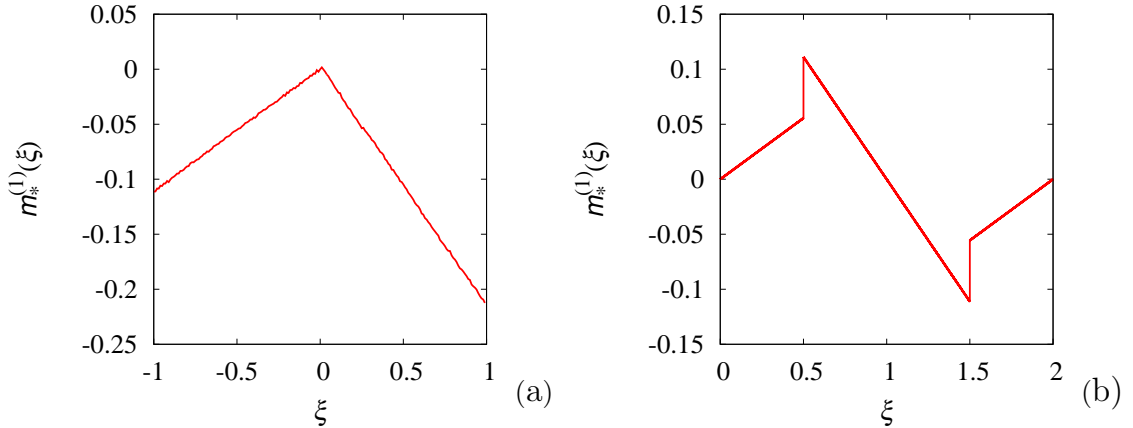}
\end{center}
\caption{Long-term spatial distribution of the first-order local moment
$m_*^{(1)}(\xi)$ within the unit periodicity cell $\xi \in (0,L)$,
$L=2$ at $D_2=D_1/4$. Panel (a) refers to simulation results
of the MuPh-LRW, with the interfacial point located at $\xi=0$,
so that $\xi=(-1,1)$. Panel (b) refers to the
analytic result eq. (\ref{eq3_10}) deriving from homogenization theory
in the Stratonovich case $\lambda=1/2$,
setting $E=0$, and defined for $\xi \in (0,2)$.
}
\label{Fig5}
\end{figure}

The validity of the Langevin-Stratonovich  model for
the long-term properties of MuPh-LRW in the case $\tau_2=\tau_1$,
finds a further confirmation in the analysis of the
stationary  first-order local
moments $m_*^{(1)}(\xi)$. Specifically,
consider a MuPh-LRW as defined and described in \cite{prl}
in the case $\delta_2=\delta_1/2=1/N$, $N=100$, $\tau_2=\tau_1$,
$L_1=L_2=L/2=1$. Figure \ref{Fig5} panel (a) shows
the stationary profile of the  local first-order moments
within the periodicity cell of the lattice, obtained
from stochastic lattice simulations involving $N_p=10^6$
particles. Since in the numerical simulation of the lattice
dynamics, the phase interface is located at $\xi=0$, the
unit periodicity cell is defined for $\xi \in (-1,1)$,
where $\xi \in (-1,0)$ corresponds to phase ``1'',
while $\xi \in (0,1)$ to phase ``2''.
Figure \ref{Fig5} panel (b) depicts the profile
of $m_*^{(1)}(\xi)$ deriving from eq.  (\ref{eq3_10}), i.e.,
from the homogenization  theory of the Langevin-Stratonovich equation,
setting the constant $E=0$. In this case, the unit periodicity
cell has been defined for $\xi \in (0,2)$, so that
$\xi=1/2$, and $\xi=3/2$ correspond to the interfacial points separating the
two phases, and $m_*^{(1)}(\xi)$ is a periodic function of $\xi$.
Apparently, the two profiles depicted in figure \ref{Fig5}
``looks different''. But this  dissimilarity
is a straightforward consequence of the
gauge associated with the long-term properties
of $m_*^{(1)}(\xi)$. As follows from eq. (\ref{eq3_10}),
$m_*^{(1)}(\xi)$ is defined modulo an irrelevant contribution
$E/\sqrt{D(\xi)}$, where $E$ is an arbitrary constant,
that does not influence the long-term dispersion properties.

\begin{figure}
\begin{center}
\includegraphics[width=12cm]{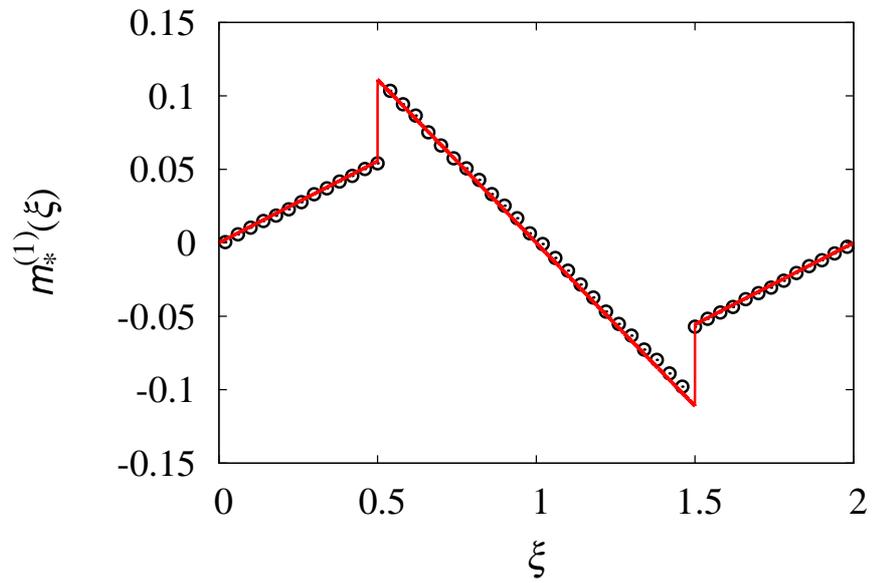}
\end{center}
\caption{Comparison of the stationary distribution of
the  first-order local moment $m_*^{(1)}(\xi)$ within the
unit periodicity cell $\xi \in (0,2)$, obtained from
stochastic simulation of MuPh-LRW (symbols $\circ$) with
the theoretical expression (\ref{eq3_10}) at $\lambda=1/2$.
The data are the same as in figure \ref{Fig5}, with the
difference that stochastic simulation data have been
referred  to the unit cell $\xi \in (0,2)$ used
in homogenization analysis upon translation, enforcing periodicity,
and to them the gauge $E/\sqrt{D(\xi)}$  has been applied with
$E=0.163$.}
\label{Fig6}
\end{figure}

Consequently, translating the lattice simulation results
onto the periodicity cell $\xi \in (0,2)$ and adding to
the simulation data the gauge $E/\sqrt{D(\xi)}$, where the
constant $E$ has been set  imposing the
condition $m_*^{(1)}(0)=0$, the profile for $m_*^{(1)}(\xi)$
derived from stochastic simulations of lattice
dynamics perfectly agrees with the theoretical
expression deriving from homogenization analysis
as depicted in figure \ref{Fig6}.
For  the sake of graphical representation, the lattice-simulation
data has been sampled with a coarser spacing than in figure \ref{Fig5}
panel (a).

\subsection{Discontinuous parabolic model: transport parameters and
equilibrium conditions}
\label{sec5_3}

Finally, let us consider the discontinuous parabolic model,
the homogenization theory of which has been addressed in Section
\ref{sec3}. An {\em a-priori} assumption of this model is
the equilibrium relation at the interfacial points separating the
two lattice phases, defined by the partition coefficient
$K$ regulating particle redistribution amongst the two phases.

From  a microscopic point of view, i.e., in terms of stochastic microdynamics,
there is no Langevin equation driven by Wiener perturbations admitting
this model
 as its Fokker-Planck equation, and that can be derived as the
limit of a smooth diffusivity profile $D(x,\varepsilon)$
in the limit for $\varepsilon \rightarrow 0$. The latter 
class of models is considered in paragraph \ref{sec3_1}, leading
to the expression (\ref{eq3_13a3}) for $D_{\rm eff}$. Moreover,
by its nature, the  discontinuous parabolic
 model contains an adjustable parameter,
given by the partition coefficient itself.

Viewed in a broader perspective, the discontinuous parabolic
model is a classical continuous transport model that
involves both equilibrium information, expressed by $K$, and
transport parameters, corresponding to the phase diffusivities
$D_h$, $h=1,2$.

An interesting property of this model stems from the following
observation. If the equilibrium partition coefficient $K$ is
chosen in order to satisfy the correct equilibrium relations occurring in 
MuPh-LRW, i.e.,
\begin{equation}
K= \frac{b_1}{b_2} = \frac{\delta_1 \, \tau_1}{\delta_2 \, \tau_2}
\label{eq5_4}
\end{equation}
then the homogenization analysis developed for it in Section \ref{sec3}
 provides
the correct expression for the effective dispersion coefficient
observed in MuPh-LRW processes.

For  the sake of simplicity, let us prove this statement for $L_1=L_2=L/2$.
Consider eq. (\ref{eq3_29}) for the effective diffusion coefficient
deriving from the discontinuous parabolic model for $L_1=L_2$, and assume
that $K$ is expressed by eq. (\ref{eq5_4}).
Eq. (\ref{eq3_29}) can be rewritten as
\begin{equation}
\frac{2}{D_{\rm eff}} = \frac{1}{2} \, (1+K) \, \frac{(D_1+K \, D_2}{K \, D_1 \, D_2} =
\frac{1}{2} \left (1+ \frac{b_1}{b_2} \right ) \, \left ( \frac{1}{K \, D_2}
+ \frac{1}{D_1} \right )
\label{eq5_5}
\end{equation}
Since $D_h= b_h^2 \, \tau_h /2$, $h=1,2$, expressing the diffusivities
in terms of the lattice velocities, eq. (\ref{eq5_5}) becomes
\begin{equation}
\frac{2}{D_{\rm eff}} = \left ( \frac{1}{b_1} + \frac{1}{b_2} \right )
\, \left ( \frac{1}{b_1 \, \tau_1} + \frac{1}{b_2 \, \tau_2} \right )
\label{eq5_6}
\end{equation}
that is exactly eq. (\ref{eq4_21}).

This result admits noteworthy implications in the parabolic/hyperbolic
setting of transport theories.
Consider the continuous description of MuPh-LRW in the
presence of ideal interfacial conditions. With reference to
lattice dynamics, interfacial points are perfectly neutral
with respect to transport and they do no add any constraints
on particle redistribution amongst the lattice phases. They acts
as  unavoidable ``passive dislocations'' in order to connect
two lattices possessing different ``space-time'' dynamic properties.
Their passive (neutral) nature implies that there are no extra physical
conditions
(and, as a consequence, no additional parameters) associated
with the local particle dynamics from-and-towards an interfacial point.

This fact is perfectly accounted for in the hyperbolic transport
model (\ref{eq2_1}), or in its smoothened version (\ref{eq4_2}),
which define the process exclusively in terms of the couple
of lattice parameters $(\delta_h,\tau_h)$ per phase or, equivalently,
of their dynamic counterparts $(b_h,\lambda_h)$. 
Particle redistribution amongst the two phases is just the consequence of the
dynamic properties characterizing the two  phases, and specifically
of the ratio of the two lattice velocities $b_1/b_2$.

In point of fact, the physical justification of  the
discontinuous parabolic model is still rooted in the
hyperbolic hydrodynamic theory of MuPh-LRW, as it is easy
to check that it represents the Kac limit of the hyperbolic
model (\ref{eq4_2}), in the case $\varepsilon \rightarrow 0$,
when $b_h = b_0 \, \widetilde{b}_h$, $\lambda_h = \lambda_0 \, \widetilde{\lambda}_h$, $h=1,2$, and the parameters $b_0$ and $\lambda_0$
diverge keeping fixed the ratio $b_0^2/2 \, \lambda_0 =1$.
In this case, the continuity conditions for the partial
fluxes, $b_1 \, p_{\pm,1} |_{x_0}= b_2 \, p_{\pm,2} |_{x_0}$,
become $J_1|_{x_0}=J_2|_{x_0}$, corresponding to the
continuity of the overall flux, and 
\begin{equation}
\left . \frac{p_2}{p_1} \right |_{x_0} = \frac{b_1}{b_2}=
\frac{\widetilde{b}_1}{\widetilde{b}_2}= K
\label{eq4_xx}
\end{equation}
defining the value of the equilibrium constant.

The discontinuous parabolic model attempts to describe lattice
dynamics using the classical parabolic approach to transport:
in the absence of biasing fields, the probability flux is
proportional to the gradient of probability density with
reverse sign. It induces the occurrence of the second-order Laplacian
contribution in the balance equation as a consequence of the
effects of random fluctuations, and the quantification of their
intensity is expressed in terms of a unique dynamic group having the
physical dimension of a squared length per unit time, thus
corresponding to a diffusion coefficient.

The space-time heterogeneity of MuPh-LRW  is defined by the couple
of parameters $(\delta_h,\tau_h)$ per phase, which act in a separate
way in order to determine the emergent macroscopic transport 
properties, such as the long-term effective dispersion in periodic
lattices. In a parabolic model, the spatial and time scales
associated with $(\delta_h,\tau_h)$ are wrapped and compressed
into the unique transport quantity $D_h = \delta_h^2/2 \, \tau_h$.
As a consequence of this,  the separation of the emergent effects
determined by the  influence of $\delta_h$ and $\tau_h$,
 clearly appearing in eq.  (\ref{eq4_21}), becomes infeasible.

It follows from the above reasoning, that 
the only way to describe a MuPh-LRW in the presence of ideal
interfacial conditions within a parabolic scheme, is to include an additional
parameter, represented by the phase partition coefficient $K$,
in order to supply for the lost information on the characteristic
lattice velocities.
 To the parameter $K$, an equilibrium interpretation
can be attributed, so that the discontinuous parabolic
model can be interpreted as resulting from the necessary
interplay between equilibrium ($K$) and non-equilibrium ($D_1,\,D_2$)
properties.

But the equilibrium explanation for the discontinuous
transport model, necessary for justifying its setting, is essentially
a ``formal superstructure'' added to it in order to compensate
for its intrinsic dynamic deficiency, associated with the impossibility
of defining a velocity parameter for the stochastic fluctuations
in each phase.

It would be interesting to explore whether a similar interpretation
of the use of equilibrium concepts within transport models could
be extended to other phenomenologies. Of course,
the present analysis of MuPh-LRW applies to ideal interfacial
conditions, where interfacial points do not exert any selective action.
Slightly different is the case of non-ideal interfaces, which
are characterized by their own local dynamics. This
issue will be discussed elsewhere, in connection with
the theory of MuPh-LRW in the presence of non-ideal interfaces.

\section{Concluding remarks}
\label{sec6}

This article has developed the homogenization theory underlying the
multiphase properties of lattice random walks outlined in \cite{prl}
in the presence of a discontinuous distribution of lattice
spacings and hopping times in two lattice phases.

Apart from providing  the necessary technical results
complementing the hyperbolic characterization of these lattice
models in a continuous setting, with specific focus on
long-time/large-distance dispersion properties,
there are some
observations of general validity that require attention
and  that can be possibly  extended to other classes of particle
systems.

The first observation is that, even in the presence of ideal
interfaces separating  the MuPh-LRW phases, there is no
parabolic model deriving from a simple stochastic description
of particle motion, expressed in the form of Langevin equations
driven by Wiener fluctuations that provides a consistent quantitative
interpretation of the long-term/large-distance results obtained in
periodic MuPh-LRW systems, over all the range of values of lattice transport
parameters.  Conversely, the hyperbolic model provides in this
case a simple and general explanation of the observed behavior.
Parabolic transport models, and specifically the Stratonovich-based
interpretation of the microscopic dynamics applies exclusively
in the case the phase heterogeneity involves exclusively
the lattice spacings, with  a uniform hopping time  characterizing
the two  phases, and the Ito-based interpretation
yields the correct dispersion coefficient when the heterogeneity
derives exclusively from a mismatch of the hopping times
in the two lattice phases.

The only way parabolic continuous models can  interpret correctly
the observed behavior of MuPh-LRW in periodic structures is  when,
{\em a-priori}, an equilibrium relation at the interfaces
between the two phases is enforced, consistently with the
partition relation deriving from the hyperbolic theory of
ideal interfaces. 

The assessment of the equilibrium conditions (for
an ideal lattice interface) is an unavoidable technical
necessity associated with the mathematical structure of parabolic
transport models, and not a physical requisite of the dynamics
of the particle system. This stems from the fact that a parabolic
transport model, when no biasing field-effect are present,
 is characterized, by its nature, by a unique transport coefficient
for each lattice phase, given by the phase diffusivity
$D_h=\delta_h^2  /2 \,\tau_h$. 

Conversely, the hyperbolic continuous model for MuPh-LRW involves 
two systems of transport parameters for each lattice phases,
$b_h=\delta_h/\tau_h$ and $\lambda_h=1/\tau_h$, decoupling the
effects of spatial and timescales involved, and providing
a correct quantitative representation of the long-term dynamics.
In point of fact, the correct predictions of the discontinuous
parabolic model for $D_{\rm eff}$ in the case the equilibrium constant $K$
is chosen equal to the ratio of the phase velocities, is 
a further support to the hyperbolic hydrodynamic description,
as the discontinuous parabolic model is the Kac limit
of the hyperbolic description, and in pure diffusion,
the long-term properties of diffusive hyperbolic dynamics
(in the absence of deterministic biasing fields)
coincide with the Kac-limit predictions \cite{gpk1}.

The analysis in this article has been focused on ideal interfacial conditions
at the separation points of the lattices  phases. The extension of
homogenization analysis to non-ideal lattice interfaces will be developed
in
forthcoming works.

\end{document}